# Automated Setup to Accurately Calibrate Electrical DC Voltage Generators

Flavio Galliana, Pier Paolo Capra, Roberto Cerri and Marco Lanzillotti

*Abstract*—At National Institute of Metrological Research (INRIM), an automated setup to calibrate DC Voltage generators, mainly top-level calibrators from 1 mV to 1 kV has been developed. The heart of the setup is an INRIM-built automated fixed ratios DC Voltage divider. The significant achievement of this setup is the possibility to interconnect the divider, a DMM characterized in linearity, a DC Voltage Standard and a DC Voltage generator under calibration and automatically to manage the calibration process. This calibration method allows to save a lot of time, to improve the reliability and to increase the accuracy of the calibration of generators. The relative uncertainties of the system span from $0.6\times10^{-6}$ to $1.2\times10^{-4}$ improving the previous capabilities of the INRIM laboratory for calibration of programmable multifunction instruments. In addition, this system allows to avoid the employment of several Standards (some of them still manually operating) carrying out the entire process without changing the setup configuration and without the presence of operators. The concept of this setup can be transferred to secondary high-level electrical calibration Laboratories that could be consider it useful for their calibration activities.

## I. INTRODUCTION

In the field of electromagnetic metrology, in DC and AC low frequency electrical measurements, programmable multifunction generating and measuring instruments as calibrators (MFCs) and multi-meters (DMMs) [1] play an important role. They are widespread in primary, secondary and industrial calibration laboratories. In fact, they combine the five electrical functions (DC and AC voltage and current and DC resistance) generating or measuring in very large measurement fields. To achieve their best performance, these instruments require periodic and complex calibration operations, often involving several calibrated instruments and Standards. Traditionally, in DC Voltage, manually operating dividers were involved [2, 3]. Consequently, the calibration process was particularly laborious requiring a lot of time increasing the calibration costs. To overcome these problems, at the National Institute of Metrological Research (INRIM), in the laboratory for calibration of programmable multifunction instruments (INRIM-Lab) a setup to calibrate DC Voltage generators allowing the interconnection of a generator, of a DMM and of a DC Voltage Standard with an INRIM-built high accuracy automated fixed ratios standard divider, has been developed (Fig. 1). An advantage of this setup is to perform the calibration of generators avoiding any manipulation or change of the circuit configuration during the calibration process. Being based on the 10 V Standard and on the DMM linearity, the calibration process performs, as first step, the calibration of the DMM at 10 V, then the temporary calibration of the 10:1 and 100:10 divider ratios using the Unit Under Test (UUT) DC Voltage generator as auxiliary voltage generator. Last step is the UUT calibration in the 1 mV–1 kV range, using the INRIM divider and the DMM as Standards. The full process takes about one day, according to the chosen delays and repetitions optimized to ensure its reliability. Nevertheless, the three steps are independent each other and can therefore be carried out at different times. The system has been developed specifically involving the DMM HP3458A [4] for its excellent linearity on the 10 V range and considering as UUTs the MFCs J. Fluke 5700 A and 5720 A [5] (block scheme of Fig. 1). Nevertheless, other generator or calibrator models can also be calibrated with setup customizing the calibration program. The full calibration process is complete in 10–12 hours while in the past for this process about a week was necessary with the constant presence of the operators introducing undesired noises.

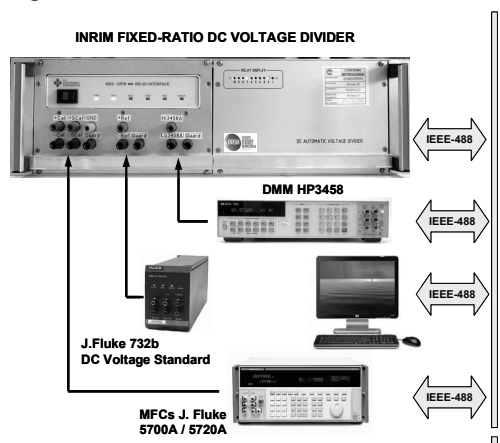

Fig.1. Block scheme of the INRIM setup to calibrate DC Voltage generators.

## II. THE INRIM-BUILT FIXED-RATIOS DC VOLTAGE DIVIDER

The INRIM-built DC Voltage resistive divider (external view

The authors are with the Dept. Innovation and Metrology Services (STALT), Strada delle Cacce, 91 -10135 Turin, Italy.
f.galliana@inrim.it

in Fig. 1) allows the 10:1, 100:10 and 100:1 division ratios and can be connected with a generator, a DMM and a DC Voltage Standard. The divider is made combining one hundred bulk metal foil 10 kΩ type Vishay resistors in three series (R, 9R and 90R (Fig. 2) selected in groups of ten (sets) [6]. These resistors sets are placed into a copper box connected to a guard circuit. The box is, in turn, placed into an aluminum case connected to the ground potential. In the same case is housed another copper box in which ten low thermal voltages bistable relays are inserted. These relays are fixed on a teflon support and are linked to a GPIB interface to allow the remote selection of the divider sets (see Fig. 3).

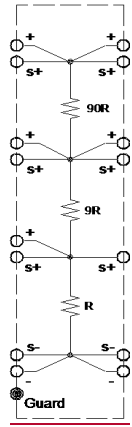

Fig. 2. Principle scheme of the DC Voltage Divider. S+ and S- are respectively the positive and negative voltage terminals (sense) while + and – are the positive and negative current terminals.

Taking advantage from the DMM excellent linearity on the 10 V range, the divider ratios 10:1 and 100:10 can be temporary calibrated. These calibrations are made by means of the measurement setup shown in Fig. 3 (only excluding the 10V Standard). In this case a UUT acts as high stability auxiliary voltage generator. The value of the 10:1 ratio is obtained applying 10 V on 10R and reading with the DMM, on the DMM 10 V range, before on 10R and after on R. The 100:10 ratio value is obtained applying 100 V before on 100R and after on 10R and reading with the DMM, on the DMM 10 V range, on R. The 100:1 ratio value is obtained from the 10:1 and 100:10 ratio values. In [6] preliminary uncertainty budgets of these temporary calibrations were made. These budgets have been now re-evaluated taking also into-account the experience in the utilization of the instrument.

TABLE I
RELATIVE STANDARD UNCERTAINTIES OF THE DIVIDER RATIOS TEMPORARY CALIBRATION.

| 10:1 ratio | | | 100:10 ratio | | |
|---|---|---|---|---|---|
| Component | type | 1 σ (×10$^{-8}$) | Component | type | 1 σ (×10$^{-8}$) |
| Noise on 10R | A | 5.0 | Noise on 10R | A | 5.0 |
| Noise on R [1] | A | 1.5 | Noise on R[1] | A | 1.5 |
| Transfer Accuracy 10 V [4] | B | 5.8 | Transfer Accuracy 10 V | B | 5.8 |
| Transfer Accuracy 1 V (10 V range) | B | 3.2 | Transfer Accuracy 1 V (10 V range) | B | 3.2 |
| 10 V resolution | B | 0.6 | 10 V resolution | B | 0.6 |
| 1 V resolution | B | 5.8 | 1 V resolution | B | 5.8 |
| Generator stability | B | 2.0 | Generator stability | B | 5.2 |
| Circuit 10 V effect | B | 1.7 | Circuit 10 V effect | B | 1.7 |
| Circuit 1 V effect | B | 2.9 | Circuit 1 V effect | B | 2.9 |
| Voltage coefficient | B | 1.2 | Voltage coefficient | B | 1.2 |
| Different load effect[2] | B | 6.0 | | | |
| **RSS** | | **17.2** | **RSS** | | **16.8** |

[1] This measurement is also made on the DMM 10 V range.

Then, for a 95 % confidence level, the relative expanded uncertainties of the temporary calibration of the 10:1 and 100:10 divider ratios are both $3.4\times10^{-7}$. The expanded relative uncertainty of the calculated 100:1 ratio has been evaluated equal to $5.9\times10^{-7}$ considering a partial correlation between the 10:1 and the 100:10 ratio values evaluations. To validate this divider temporary calibration method, the divider itself is periodically calibrated, at the INRIM-Laboratory for DC Voltage ratios calibration, vs. the DC Voltage-Ratio standard divider Datron 4902S which was employed as travelling standard in the CCEM-K8 [7] international comparison. Fig. 4 shows the results of the calibrations of the INRIM fixed ratios-divider with both methods in 2016. As the measurements agree, the method of the temporary calibration of the fixed-ratios divider can be considered reliable.

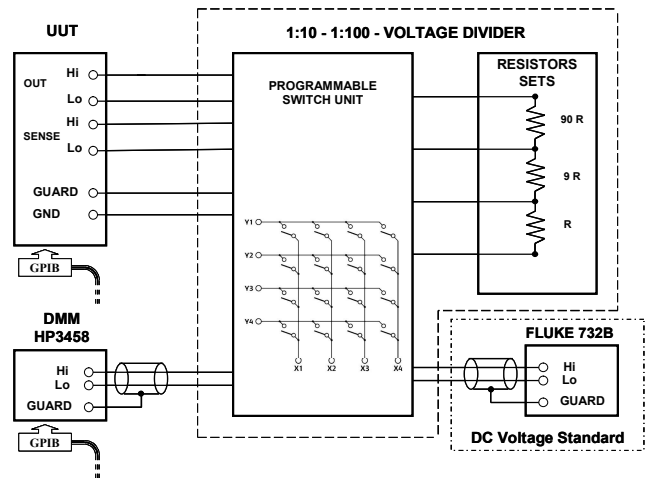

Fig. 3. Complete setup of the system to calibrate DC Voltage generators: the divider can be connected to the generator, to the DMM and to the DC Voltage Standard.

[2] The input impedance of the available DMMs of the same model were accurately measured and the item with the higher impedance (about $8.6\times10^{11}$ Ω) was enrolled in the measurement setup This component has been evaluated considering this DMM. For the 100:10 ratio this component is negligible as the DMM in both case reads on R.

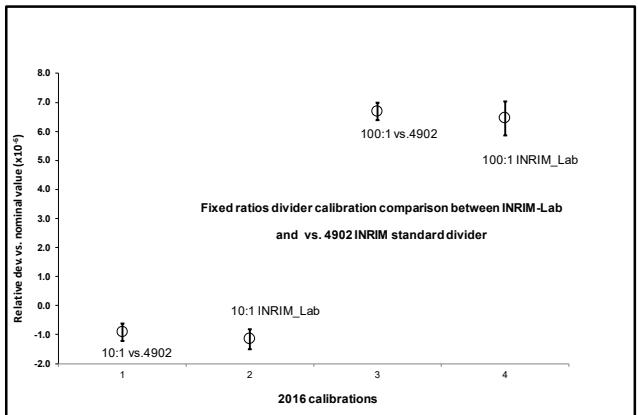

Fig. 4. INRIM fixed ratios divider calibration comparison between the INRIM-Lab vs. the INRIM standard DC Voltage ratio divider. The uncertainty bars correspond to a 95% confidence level.

### III. THE AUTOMATIC CALIBRATION SETUP OF DC VOLTAGE GENERATORS

The first step of the complete calibration process of DC Voltage generators, consisting in the calibration of the 10 V value of the DMM vs. a 10 V Standard, calibrated in turn vs. the Josephson effect, is made performing series of positive and negative readings with the DMM connected to the DC Voltage Standard and successively performing a series of readings with the DMM input leads shorted to the low voltage terminal of the Voltage Standard. The average of the positive and negative means corrected by the offsets and by the deviation from the 10 V reference value, are considered and stored. The linearity of the 10 V range of the high precision DMM involved in the calibration setup is periodically verified vs. the Josephson effect [9–11]. The verification of the DMM linearity and the fast calibration process allow to consider, in the uncertainties evaluation, the Transfer Accuracy/linearity specifications of the DMM which are significantly smaller than the accuracy specifications. The second step of the calibration process is the temporary calibration of the 10:1 and 100:10 divider ratios and the calculation of the 100:1 ratio value. The final step is the calibration of generators from 1 mV to 1 kV, using the INRIM divider and the DMM as standards. This step consists, in turn, of several steps shown in Fig. 7. Main ones are:
- Calibration of the positive and negative decade values 100 mV, 1 V and 10 V (starting from 10 V with a step-down process) and the 100 V and 1 kV values with a step-up process;
- Calibration of not-decade, positive and negative values from 1mV to 9 V with a step-down process;
- Calibration of not-decade, positive and negative values from 20 V to 800 V with a step-up process.

For each measurement point the average of the positive and negative mean values provided by the UUT or read by the DMM and their relative deviation from the reference values are considered and stored.

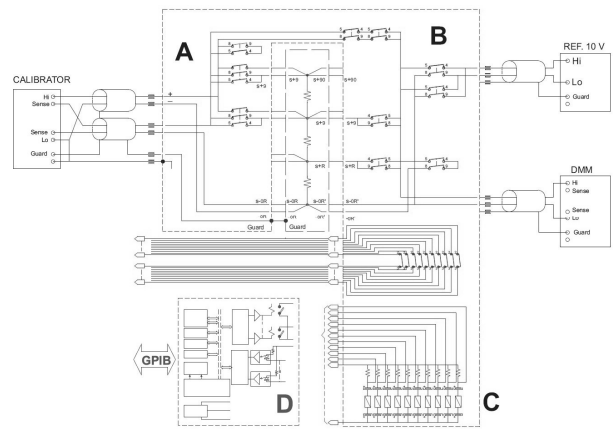

Fig. 5 – The complete relays system of the divider. The sets A) and B) are designed to invert both the voltage of the calibrator (generator) and of the 10 V reference voltage. The last relays group controls the resistors sets of the fixed-ratio divider. C) the relays coils are pulse controlled through two electrolytic capacitors, to decouple the relays from the digital control circuit. D) Commercially available GPIB to digital I/O interface.

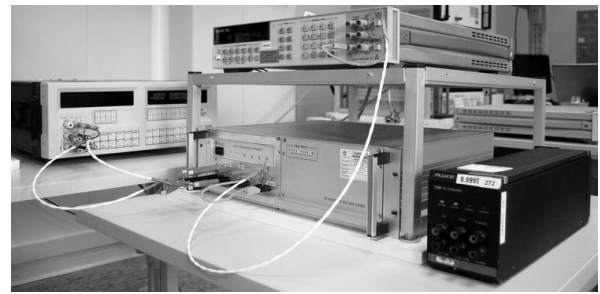

Fig. 6. Photo of the measurement setup to calibrate a top-level calibrator.

### IV. MAIN CALIBRATION POINTS OF GENERATORS

Normally at the INRIM-Lab, the calibration points chosen for the generators, mainly calibrators, are chosen referring to the Guide [12]. This guide suggests that the calibration points should be the full scale and the adjustment points of the multifunction electrical instruments. In the following, the calibration steps for some measurement values and ranges of generators are discussed. The calibration, for example of a calibrator, is always performed connecting it to the DMM through the divider. This connection allows the automation of the process.

#### A. Calibration at 10 V

The calibration of a generator, at 10V is simply performed connecting it to the DMM through the divider set to the 1:1 ratio and reading with the DMM.

#### B. Calibration at 1 V and 100 mV

The calibrations of a generator at 1 V and 100 mV are similar. The calibration at 1 V is made in two steps: In the first one, the generator is connected to the DMM operating on the 1V range through the divider set in the 10:1 ratio. The generator provides the previously calibrated 10 V. The output value from the divider, read by the DMM, allows the temporary calibration of the DMM at 1 V. In the second step, without removing the connections and setting the divider in the 1:1 ratio, the generator is set to 1 V (on the 1 V range) and the DMM directly measures this provided value calibrating the generator at 1 V. The calibration at 100 mV is like that of the 1 V, setting only the divider in the 100:1 ratio in the first step.

*C. Calibration at 100 V and 1000 V*

The calibrations of a generator at 100 V and 1000V are performed in a single step with the generator supplying alternatively these voltages and setting the divider ratio at the 10:1 and 100:1 respectively and reading with the DMM on the 10 V range.

*D. Calibration of the values from 2 V to 9 V*

The values from 2 to 9 V lie between two measurement ranges of the generator (2.2V and 11V) that are automatically selected by the measurements control program. A zero check is made at every range change. The calibration is made setting the divider in the 1:1 ratio and taking advantage of the DMM linearity. The calibration is made directly reading with the DMM the voltages provided by the generator.

*E. Calibration of the values from 200 mV to 500 mV*

The values between 200 mV and 500 mV lie also between two measurement ranges of the generator (220 mV and 2.2V). Zero check is always made at every range change. The calibration is carried out in two steps. During the first one, the divider is set to the 10:1 ratio and the generator provides the already calibrated 2 V÷5 V. The output values from the divider are read by the DMM, temporary calibrating it from 200 mV to 500 mV. In the second step, the divider is set to the 1:1 ratio and the DMM directly reads the values from 200 mV to 500 mV provided by the generator calibrating the generator in those values.

*F. Calibration of the values from 10 mV to 50 mV*

In this case the procedure utilizes the already calibrated 1 V÷5 V values (automatically changing the range of the generator from 2.2 V to 22 V). The calibration steps are the same of the range from 200 mV to 500 mV setting only the divider in the 100:1 ratio in the first step.

*G. Calibration of the values from 1 mV to 5 mV*

This procedure is like that of the point *F* with the only difference that the generator utilizes the already calibrated 100 mV ÷ 500 mV values (automatically changing the range of the generator from 220 mV to 2.2 V).

*H. Calibration of the values from 20 V to 80 V and from 200 V to 800 V*

The calibration of the generator from 20 V to 80 V is made setting the divider in the 10:1 ratio and reading the value provided by the generator directly with the DMM on the 10V range. The calibration from 200 V to 800V is the same, only setting the divider in the 100:1 ratio.

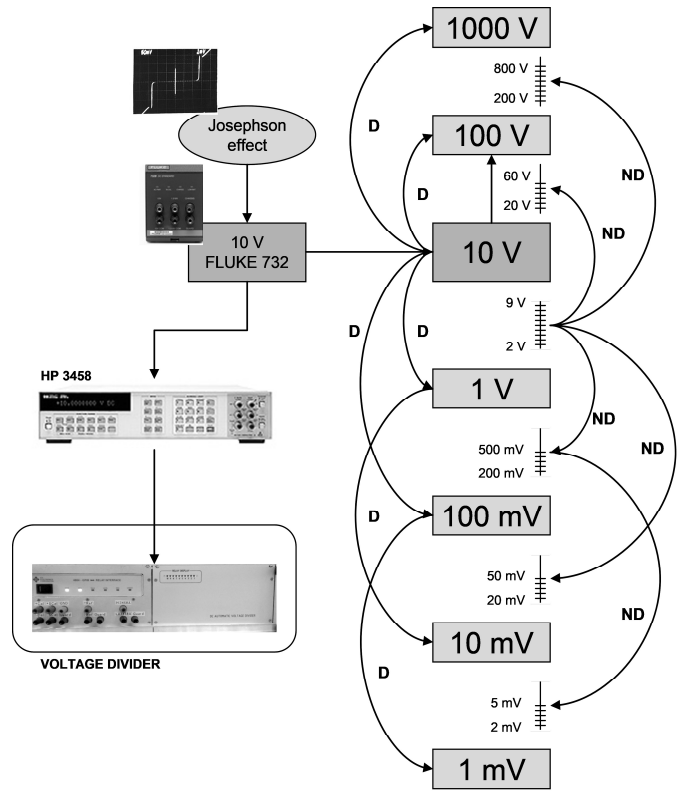

Fig. 7. Diagram of the calibration process. The value of the 10V is the key point of the traceability that is provided by a high precision Zener diode-based DC Voltage Standard directly calibrated vs. the Josephson effect. The various process steps are symbolized with D for the calibration of decade values and with ND for the calibration of not-decade values.

## V. CALIBRATION UNCERTAINTIES

In Table 1 relative standard uncertainties budget of the calibration of the DMM vs. at 10 V vs. the DC Voltage Standard is given.

TABLE II
RELATIVE STANDARD UNCERTAINTY BUDGET OF THE CALIBRATION OF THE DMM AT 10 V.

| Component | type | 1 σ ($\times 10^{-7}$) |
|---|---|---|
| DC Voltage Standard calibration | B | 2.9 |
| DC Voltage Standard temp. effect | B | 0.1 |
| DC Voltage Standard drift | B | 1.0 |
| DC Voltage Standard emfs | B | 0.1 |
| DC Voltage Standard noise at the output terminals | B | 0.3 |
| DMM noise | A | 0.3 |
| **RSS** | | **3.1** |

Then for a 95 % confidence level the relative expanded uncertainty of the calibration of the DMM at 10 V is $6.2 \times 10^{-7}$.

*A. Generators calibration uncertainties*

In Tables from 3 to 12 relative standard uncertainties budgets of the calibration of a generator at some voltages are given.

TABLE III
RELATIVE STANDARD UNCERTAINTY BUDGET OF THE CALIBRATION OF A GENERATOR AT 10 V.

| Component | type | 1 σ ($\times 10^{-7}$) |
|---|---|---|
| Circuit 10 V effect (divider 1:1) | B | 0.2 |
| DMM noise | A | 0.3 |

| Component | type | 1 σ (×10⁻⁷) |
|---|---|---|
| DMM calibration at 10 V | B | 3.1 |
| Transfer Accuracy 10 V | B | 0.6 |
| **RSS** | | **3.2** |

The relative expanded uncertainty of the calibration of a generator at 10 V is 6.3×10⁻⁷. The calibration of a generator at 1 V is made in two steps:

TABLE IV a)
RELATIVE STANDARD UNCERTAINTY BUDGET OF THE TEMPORARY CALIBRATION OF THE DMM AT 1 V (FIRST STEP).

| Component | type | 1 σ (×10⁻⁷) |
|---|---|---|
| Generator calibrated at 10 V | B | 3.2 |
| Divider 10 :1 | B | 2.0 |
| DMM noise | A | 0.3 |
| emf | B | 0.3 |
| **RSS** | | **3.8** |

TABLE IV b)
RELATIVE STANDARD UNCERTAINTY BUDGET OF THE CALIBRATION OF A GENERATOR AT 1 V (SECOND STEP).

| Component | type | 1 σ (×10⁻⁷) |
|---|---|---|
| Circuit 10 V effect (divider 1:1) | B | 0.2 |
| DMM noise | A | 0.4 |
| DMM calibration at 1 V | B | 3.8 |
| Transfer Accuracy | B | 2.3 |
| emf | B | 0.3 |
| **RSS** | | **4.4** |

The relative expanded uncertainty of the calibration of a generator at 1 V is 8.9×10⁻⁷.

TABLE V
RELATIVE STANDARD UNCERTAINTY BUDGET OF THE CALIBRATION OF A GENERATOR AT 100 V.

| Component | type | 1 σ (×10⁻⁷) |
|---|---|---|
| Circuit 10 V effect (divider 10:1) | B | 2.0 |
| DMM noise | A | 0.3 |
| DMM calibration at 10 V | B | 3.1 |
| Transfer Accuracy | B | 0.6 |
| **RSS** | | **3.7** |

Then the relative expanded uncertainty of the calibration of a generator at 100 V is 7.4×10⁻⁷.

TABLE VI
RELATIVE STANDARD UNCERTAINTY BUDGET OF THE CALIBRATION OF A GENERATOR AT 5 V.

| Component | type | 1 σ (×10⁻⁷) |
|---|---|---|
| Circuit 10 V effect (divider 1:1) | B | 0.2 |
| DMM noise | A | 0.5 |
| DMM calibration 10 V | B | 3.1 |
| emf | B | 0.1 |
| Transfer Accuracy | B | 0.9 |
| Deviation from linearity at 5 V[3] | B | 0.6 |
| **RSS** | | **3.6** |

So, the relative expanded uncertainty of the calibration of a generator at 5 V is 7.2×10⁻⁷. The calibration of a generator at 0.5 V is made in two steps:

TABLE VII a)

[3] This value is obtained from the calibration certificate concerning the DMM linearity verification on the 10 V range.

TABLE VII a)
RELATIVE STANDARD UNCERTAINTY BUDGET OF THE TEMPORARY CALIBRATION OF THE DMM AT 0.5 V (FIRST STEP).

| Component | type | 1 σ (×10⁻⁷) |
|---|---|---|
| Generator calibrated at 5 V | B | 3.6 |
| Divider 10 :1 | B | 2.0 |
| DMM noise | A | 1.0 |
| Emf | B | 0.6 |
| Term. Eff. on generator | B | 0.9 |
| **RSS** | | **4.4** |

TABLE VII b)
RELATIVE STANDARD UNCERTAINTY BUDGET OF THE CALIBRATION OF A GENERATOR AT 0.5 V (SECOND STEP).

| Component | type | 1 σ (×10⁻⁷) |
|---|---|---|
| Circuit 10 V effect (divider 1:1) | B | 0.2 |
| DMM noise | A | 1.0 |
| DMM calibration 0.5 V | B | 4.4 |
| Transfer Accuracy | B | 2.3 |
| emf | B | 0.6 |
| **RSS** | | **5.1** |

The relative expanded uncertainty of the calibration of a generator at 0.5 V is 1.0×10⁻⁶. Also the calibrations of a generator at 50 mV and 5 mV are made in two steps:

TABLE VIII a)
RELATIVE STANDARD UNCERTAINTY BUDGET OF THE TEMPORARY CALIBRATION OF THE DMM AT 50 mV (FIRST STEP).

| Component | type | 1 σ (×10⁻⁷) |
|---|---|---|
| Generator calibrated at 5 V | B | 3.6 |
| Divider 100 :1 | B | 3.4 |
| DMM noise | A | 10 |
| emf | B | 12 |
| Term. Eff. on generator | B | 0.9 |
| **RSS** | | **16** |

TABLE VIII b)
RELATIVE STANDARD UNCERTAINTY BUDGET OF THE CALIBRATION OF A GENERATOR AT 50 mV (SECOND STEP).

| Component | type | 1 σ (×10⁻⁷) |
|---|---|---|
| Circuit 10 V effect (divider 1:1) | B | 0.2 |
| DMM noise | A | 10 |
| DMM calibration | B | 16 |
| Transfer Accuracy | B | 0.6 |
| emf | B | 12 |
| **RSS** | | **23** |

The relative expanded uncertainty of the calibration of a generator at 50 mV is 4.6×10⁻⁶.

TABLE IX a)
RELATIVE STANDARD UNCERTAINTY BUDGET OF THE TEMPORARY CALIBRATION OF THE DMM AT 5 mV (FIRST STEP).

| Component | type | 1 σ (×10⁻⁶) |
|---|---|---|
| Generator calibrated at 0.5 V | B | 0.5 |
| Divider 100 :1 | B | 0.3 |
| DMM noise | A | 8.0 |
| emf | B | 1.7 |
| Term. Eff. on generator | B | 0.1 |
| **RSS** | | **8.2** |

TABLE IX b)
RELATIVE STANDARD UNCERTAINTY BUDGET OF THE CALIBRATION OF A GENERATOR AT 5 mV (SECOND STEP).

| Component | type | 1 σ (×10⁻⁶) |
|---|---|---|
| Circuit 10 V effect (divider 1:1) | B | 0.02 |

| Component | type | 1 σ (×10⁻⁷) |
|---|---|---|
| DMM noise | A | 8.0 |
| DMM calibration | B | 8.2 |
| Transfer Accuracy 10 V | B | 0.6 |
| emf | B | 1.7 |
| **RSS** | | **11.6** |

The relative expanded uncertainty of the calibration of the generator at 5 mV is $2.3 \times 10^{-5}$.

TABLE X
RELATIVE STANDARD UNCERTAINTY BUDGET OF THE CALIBRATION OF A GENERATOR AT 50 V.

| Component | type | 1 σ (×10⁻⁷) |
|---|---|---|
| Circuit 10 V effect (divider 10:1) | B | 2.0 |
| DMM noise | A | 0.6 |
| DMM calibration at 10 V | B | 3.1 |
| DMM Transfer Accuracy | B | 0.9 |
| Deviation from linearity at 5 V | B | 0.6 |
| **RSS** | | **4.2** |

The relative expanded uncertainty of the calibration of a generator at 50 V is then $8.42 \times 10^{-7}$.

TABLE XI
RELATIVE STANDARD UNCERTAINTY BUDGET OF THE CALIBRATION OF A GENERATOR AT 500 V.

| Component | type | 1 σ (×10⁻⁷) |
|---|---|---|
| Circuit 10 V effect (divider 100:1) | B | 3.4 |
| DMM noise | A | 0.4 |
| DMM calibration at 10 V | B | 3.1 |
| Transfer Accuracy | B | 0.9 |
| emf | B | 0.2 |
| Deviation from linearity at 5 V | B | 0.6 |
| **RSS** | | **4.9** |

So the relative expanded uncertainty of the calibration of a generator at 500 V is $9.8 \times 10^{-7}$.

TABLE XII
RELATIVE STANDARD UNCERTAINTY BUDGET OF THE CALIBRATION OF A GENERATOR AT 1000 V.

| Component | type | 1 σ (×10⁻⁷) |
|---|---|---|
| Circuit 10 V effect (divider 100:1) | B | 3.4 |
| DMM noise | A | 0.3 |
| DMM calibration at 10 V | B | 3.1 |
| Transfer Accuracy | B | 0.6 |
| **RSS** | | **4.6** |

The relative expanded uncertainty of the calibration of a generator at 1000 V is $0.9 \times 10^{-6}$. In Table 13 the relative expanded uncertainties (i.e. the best measurement capabilities) of the INRIM-Lab for the calibration of generators from 1 mV to 1 kV before and after the introduction of this setup, are listed.

TABLE XIII
RELATIVE EXPANDED UNCERTAINTIES OF THE CALIBRATION OF GENERATORS FROM 1 mV TO 1 kV. THE TABLE IS REFERRED TO THE CALIBRATION OF A J. FLUKE 5700A CALIBRATOR. THE UNCERTAINTIES ARE GROUPED IN MEASUREMENT RANGES ACCORDING TO THE PUBLICATION CRITERIA OF THE CIPM-MRA DATABASE[4].

| DC Voltage Ranges | Previous INRIM-Lab capabilities (µV/V)[5] | INRIM-Lab capabilities with the setup (µV/V)[4] |
|---|---|---|
| ± 1 mV | 192 | 118 |
| ±1 mV ÷ ± 3 mV | 65 | 43 |
| ± 3 mV ÷ ± 5 mV | 42 | 23 |
| ± 5 mV ÷ ± 10 mV | 23 | 12.4 |
| ± 10 mV ÷ ± 30 mV | 11 | 4.6 |
| ± 30 mV ÷ ± 50 mV | 8.5 | 4.6 |
| ± 50 mV ÷ ± 200 mV | 2.2 | 1.4 |
| ± 200 mV ÷ ± 300 mV | 1.9 | 1.3 |
| ± 300 mV ÷ ± 500 mV | 1.7 | 1.0 |
| ± 500 mV ÷ ± 1 V | 0.9 | 0.9 |
| ± 1 V ÷ ± 2 V | 1.2 | 0.9 |
| ± 2 V ÷ ± 3 V | 0.9 | 0.8 |
| ± 3 V ÷ ± 4 V | 0.8 | 0.8 |
| ± 4 V ÷ ± 5 V | 0.8 | 0.7 |
| ± 5 V ÷ ± 6 V | 0.7 | 0.7 |
| ± 6 V ÷ ± 7 V | 0.7 | 0.7 |
| ± 7 V ÷ ± 8 V | 0.7 | 0.7 |
| ± 8 V ÷ ± 9 V | 0.7 | 0.7 |
| ± 9 V ÷ ± 10 V | 0.6 | 0.6 |
| 10 V ÷ ± 12 V | 0.8 | |
| ± 12 V ÷ ± 20 V | 0.8 | 1.0 |
| ± 20 V ÷ ± 30 V | 0.9 | 0.9 |
| ± 30 V ÷ ± 40 V | 1.1 | 0.9 |
| ± 40 V ÷ ± 50 V | 1.2 | 0.8 |
| ± 50 V ÷ ± 60 V | 1.4 | 0.8 |
| ± 60 V ÷ ± 80 V | 1.7 | 0.8 |
| ± 80 V ÷ ± 100 V | 0.8 | 0.7 |
| ± 100 V ÷ ± 200 V | 1.6 | 1.1 |
| ± 200 V ÷ ± 300 V | 1.4 | 1.0 |
| ± 300 V ÷ ± 400 V | 1.4 | 1.0 |
| ± 400 V ÷ ± 500 V | 1.4 | 1.0 |
| ± 500 V ÷ ± 600 V | 1.3 | 0.9 |
| ± 600 V ÷ ± 800 V | 1.3 | 1.0 |
| ± 800 V ÷ ± 1000 V | 1.3 | 0.9 |

## VI. MEASUREMENT RESULTS

Fig. 8 shows the results of the calibration of a MFC at some DC voltage positive and negative values utilizing the described measurement setup. These values are the readings of the DMM at the output of the INRIM fixed-ratios divider. Calibration values are obtained averaging the DMM acquired measurements at reached suitable readings stabilization.

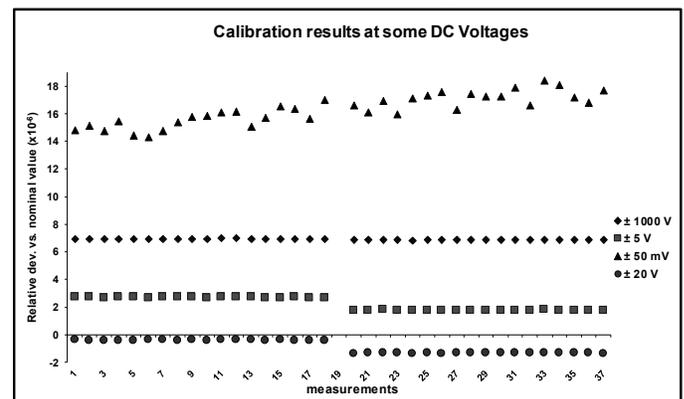

Fig. 8. Calibration results at typical DC Voltage values of a top-level calibrator.

---

[4] The CIPM Mutual Recognition Arrangement (CIPM MRA) is the framework through which National Metrology Institutes demonstrate the international equivalence of their measurement standards and the calibration and measurement certificates they issue. The outcomes of the Arrangement are the internationally recognized (peer-reviewed and approved) Calibration and Measurement Capabilities (CMCs) of the participating institutes.

[5] For same measurement values in consecutive ranges (as the last value of a range and the first of the following range, excluding the 1 mV) the better uncertainty value should be considered.

## VII. Conclusions

From Table 13 it is possible to see that the relative uncertainties (best measurement capabilities) of the INRIM-Lab for the calibration of DC Voltage generators from 1 mV to 1 kV, after the introduction of the new setup, significantly improved. Another achievement with the new system is the time saving and the measurement reliability as the system can operate without the presence of operators for example during the week-ends. The DC Voltage results of a measurement comparison between the INRIM-Lab and a high level secondary electrical Laboratory carried out with the new system [13] were re-evaluated[6]. The agreement with the measurements of this laboratory is still maintained even considering the new uncertainties. This exercise represents a first validation of the new setup. In future, further comparisons to check the reliability of the system will be made, for example the calibration of the INRIM-built selectable-value transportable high dc voltage standard (THVS), comparing the results obtained with the setup with those obtained with the THVS calibration system [14]. Other aims will be the updating the INRIM capabilities in the CIPM MRA database and the transfer of the setup concept to secondary electrical calibration Laboratories or to other National Metrology institutes.

## Acknowledgment
## Acknowledgment

The authors wish to thank Cristina Cassiago that wrote the software to manage the calibration process and for her support in the elaboration of the paper.

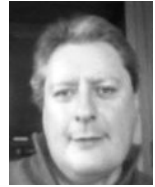

**Flavio Galliana** was born in Pinerolo, Italy, in1966. He received the M.S. degree in physics from the Università degli Studi di Torino, Torino, Italy, in 1991.In 1993 he joined the Istituto Elettrotecnico Nazionale "Galileo Ferraris (IEN), Torino, where he was involved in precision high resistance measurements. He also joined the "Accreditation of Laboratories" Department of IEN. From 2001 to 2005 he was responsible of the Accreditation of Laboratories" Department of IEN. Since 2006, being IEN part of the National Institute of Metrological Research (INRIM), he was involved in precision resistance measurements and recently in ILCs technical managements.

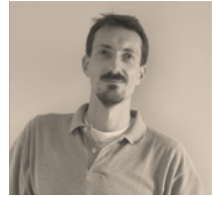

**Pier Paolo Capra** was born in Torino, Italy, in 1965. He received the Technical High-School degree in chemistry from Istituto Tecnico Industriale Statale L. Casale, Torino, in 1984, the M.S. degree in physics from the Università degli Studi di Torino, in 1996 and the Ph.D. degree in metrology from the Politecnico diTorino. In 1987, he joined the IEN, where he was involved in dc voltage and resistance precision measurement. In 1997, involved in the realization and maintenance of the national standard of dc electrical resistance at INRIM. Researcher in 2001 and is currently responsible for the national resistance standard and for the dissemination of the resistance unit.

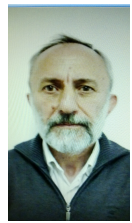

**Roberto Cerri** was born in Settimo T.se (TO), 1956. Technical Maturity, Turin. From 1994 to IEN he deals with the Josephson effect and DC voltage scale, and Laboratory accreditation assessor. With INRIM continues the activity of electrical measurements and deals with quality management.

**Marco Lanzillotti** was born in Turin, Italy, in 1981. He received the technical school degree in electronic and telecommunications from I.T.I.S. "E. Majorana", Grugliasco, Torino, in 2001. In 2002 He joined the electrical Metrology Department of IEN now I.N.RI.M., Turin, in 2002. He is involved in researches about the metrological characteristics of the high precision multifunction instruments and the ac-dc transfer in voltage e current and for the calibration of such instruments. Since 2010 he has also been involved in the activity for accreditation and auditing.

---

[6] For this comparison the old uncertainties, as still published in the BIPM-MRA CMC database, were declared because the comparison results were reported in an official document (an INRIM inter-laboratory comparison report) to be presented by the Laboratory itself to the Accreditation body for the Laboratory accreditation maintenance.